# Quantitative Measurements of Nanoscale Permittivity and Conductivity Using Tuning-fork-based Microwave Impedance Microscopy

Xiaoyu Wu[1], Zhenqi Hao[2], Di Wu[1], Lu Zheng[1], Zhanzhi Jiang[1], Vishal Ganesan[1], Yayu Wang[2,3], Keji Lai[1]

[1] Department of Physics, University of Texas at Austin, Austin TX 78712, USA

[2] State Key Laboratory of Low Dimensional Quantum Physics, Department of Physics, Tsinghua University, Bejing 100084, China

[3] Collaborative Innovation Center of Quantum Matter, Beijing 100084, China

## Abstract

We report quantitative measurements of nanoscale permittivity and conductivity using tuning-fork (TF) based microwave impedance microscopy (MIM). The system is operated under the driving amplitude modulation mode, which ensures satisfactory feedback stability on samples with rough surfaces. The demodulated MIM signals on a series of bulk dielectrics are in good agreement with results simulated by finite-element analysis. Using the TF-MIM, we have visualized the evolution of nanoscale conductance on back-gated $MoS_2$ field effect transistors and the results are consistent with the transport data. Our work suggests that quantitative analysis of mesoscopic electrical properties can be achieved by near-field microwave imaging with small distance modulation.

## I. Introduction

Near-field microwave microscopy is a rapidly evolving technique that can spatially resolve material properties at a length scale far below the freespace wavelength of microwave radiation[1,2]. Early implementations of the microscope utilized a small aperture at a microwave cavity[3,4] or a needle-like probe coupled to a microwave resonator[5-8]. Because of the lack of tip-sample distance control, these designs are intrinsically susceptible to tip damage and difficult for applications in nanoscale quantitative measurements. In the past decade, research in this field has taken advantage of the standard atomic-force microscopy (AFM) feedback control[9-12]. The AFM-compatible probes, either shielded cantilevers with a pyramidal tip[9,10] or unshielded cantilevers with an etched high-aspect-ratio metal tip[11,12], are connected to customized electronics[9,10] or vector network analyzers[11,12] through an impedance-match section. The systems demonstrate very high sensitivity (down to sub-aF at 1 – 20 GHz) in impedance detection, thus termed microwave impedance microscopy (MIM)[9,13], and good spatial resolution down to 10 – 100 nm. As a result, the MIM and other similar tools have found tremendous applications in condensed matter physics[14-17], material science[18-20], photovoltaics[21-23], device engineering[24,25], and biological science[26,27], with an ever-increasing list in the near future.

Along with the scientific exploration, the MIM technique itself has also undergone continuous improvements over the years. Most MIM experiments to date are based on contact-mode operation[9,10], during which the tip wearing is inevitable. Since the signal level is strongly affected by the condition of the tip apex, quantification of the contact-mode MIM is very difficult and extensive calibration process is needed throughout the measurements. The issue of tip degradation can be partially solved by using tapping-mode MIM[28], although the spatial resolution is usually compromised by the large dithering amplitude (50 – 100 nm) in cantilever-based systems. In this regard, the recently developed tuning-fork (TF) based MIM[29-31] with etched metal tips provides an elegant solution to the problem. Quartz TFs with small vibration amplitudes (< 10 nm) are widely used as the feedback elements in scanning probe microscopy[32,33], especially under cryogenic environments. In addition to the self-sensing capability that preserves the tip condition, TF-MIM with signal modulation also introduces other advantages. For instance, the MIM electronics in contact mode are only sensitive to the relative electrical contrast between the material of interest and a background region[13]. In the TF-MIM mode, the distance modulation automatically provides such a contrast mechanism and the demodulated MIM AC signals carry absolute information of the sample at every point[30]. In this paper, we further develop the TF-MIM by using the driving amplitude modulation (DAM) mode[34], which offers satisfactory stability on samples with rough surfaces. The demodulated MIM AC signals are analyzed by a combination of finite element analysis (FEA) of the tip-sample admittance and Fourier transformation of the real-time signals. Quantitative agreement between experimental data and simulated results can be achieved on bulk dielectrics, standard calibration samples, and field-effect transistor nano-devices. Our work suggests that the TF-MIM is an excellent tool for quantitative nanoscale imaging of electrical properties in functional materials.

## II. Experimental Setup and Analytical Methods

Fig. 1 illustrates the TF-MIM setup with DAM feedback control[34]. Conventional frequency-modulation (FM) tuning-fork AFMs using stiff cantilevers and small amplitudes prove to be a powerful tool to achieve atomic resolution in ultrahigh vacuum environments[33]. However, many samples in MIM studies may have a surface roughness of 10 to 100 nm. As the change of tip-sample distance becomes comparable to the vibration amplitude, the atomic force can jump frequently between attractive and repulsive regimes, leading to non-monotonic frequency shift throughout the experiment and instability in the feedback control. On the other hand, the power dissipation of the TF sensor depends monotonically on the tip height and can therefore be used for robust feedback control for topographic sensing on rough surfaces[35].

In our implementation, a sinusoidal voltage $V_{\text{drive}}$ at a frequency of $f_{\text{TF}}$ is generated by the oscillator to drive the TF. The TF signal is detected by a current-to-voltage (I/V) amplifier and demodulated by the HF2LI (Zurich Instruments) lock-in amplifier, whose output consists of the phase shift $\phi$ and the mechanical oscillation amplitude $A$. A built-in phase locked loop (PLL) module is employed to maintain the phase shift to a set point of $\phi_{\text{sp}}$ by changing the driving frequency, which keeps the TF on resonance. The oscillation amplitude is kept at a constant $A_{\text{sp}}$ by using a built-in proportional-integral-derivative (PID) controller. In the DAM mode, this PID output, which represents the energy dissipation due to atomic-force interaction, is fed into the z-controller of a commercial AFM (Park XE-70) as the feedback signal to maintain a constant average tip-sample distance. More details of the DAM mode and comparisons with the FM mode can be found in Ref. 34. The distance modulation also leads to the periodic change of MIM signals at the TF frequency. As shown in Fig. 1, the MIM-Im/Re signals (proportional to the imaginary and real parts of the tip-sample admittance, respectively) are demodulated at $f_{TF}$ to form the MIM-Im/Re AC outputs of the system.

Following the same recipe in Ref. 36 and 37, we electrochemically etch a W or Pt/Ir wire with a diameter of 25 µm, as seen in Fig. 2a. The wire is then glued to one prong of a quartz tuning fork and soldered to the exposed center conductor of a coaxial cable connected to the MIM circuit. The sharpness of the tip can be accurately controlled by the etching condition. In this paper, we choose a relatively blunt tip to enhance the signal strength, whereas sharper tips are preferred for high-resolution imaging. The tip-sample admittance is computed by commercial FEA software COMSOL 4.4. As shown in Fig. 2b, the tip diameter $d = 300$ nm and the half-cone angle $\theta = 6°$ are measured from the scanning electron microscopy (SEM) image in the inset of Fig. 2a. As discussed below, the only fitting parameter of the tip geometry is the radius of curvature $r$ (in this case $r \sim 600$ nm) at the apex, which can be determined by measurements on bulk dielectrics. With a good TF feedback control, we have confirmed that the tip condition can be preserved over an extended period of experiments.

At the MIM working frequency near 1 GHz, the metal wire behaves electrically as a lumped element with effective resistance $R_{tip} = 1\ \Omega$, capacitance $C_{tip} = 0.16$ pF, and inductance $L_{tip} = 10$ nH connected in series. The impedance ($Z$) match section[13] (inset of Fig. 2c) consists of a flexible quarter-wave cable (Astro-Boa-Flex III, Astrolab Inc.) of 4.8 cm and a semi-rigid tuning stub (UT-085C, Micro-Coax Inc.) of 6.0 cm. As shown in Fig. 2c, the measured reflection coefficient $S_{11}$ can be precisely reproduced by transmission-line analysis[13]. With a small load of tip-sample admittance $Y_{t-s}$, the same modeling can also yield the change of $S_{11}$ as a function of frequency. Taking into account of the input microwave power (~ -20 dBm) and the electronic gain (~ 90 dB), we can plot the conversion factor between the admittance input (in unit of nS) and the MIM output (in unit of mV) in Fig. 2c, which peaks at the working frequency of 958 MHz. Using this parameter, $Y_{t-s}$ simulated by the FEA can be directly converted to the MIM signals in a quantitative manner.

Fig. 2d shows the simulated $Y_{t-s}$ as a function of the tip-sample distance based on the tip geometry in Fig. 2b. As the tip oscillates between the contact point and a maximum height of $A_{p-p}$, the tip-sample admittance oscillates accordingly within a range of $\Delta Y$ in the approach curve. In this work, $A_{p-p}$ is kept at a relatively large value of 10 nm for a better tracking of sample surfaces. In Fig. 2e, the time dependence of $Y_{t-s}$ is calculated by correlating the tip-sample distance with the simulated admittance at each moment. In the experiment, the demodulated signals at the fundamental TF frequency, i.e., the first harmonic peak in the corresponding Fourier spectrum (Fig. 2f), are used for the MIM AC output. Since the approach curve strongly depends on the local permittivity and conductivity of the sample, the TF-MIM can quantitatively determine these electrical properties after a proper calibration.

### III. Results and Discussions

Similar to other quantitative microwave microscopy work[38,39], our first set of experiment is to calibrate the instrument with various bulk materials. For each sample, an area of 2 µm$^2$ is scanned by TF-MIM and the corresponding MIM-Im AC signal is averaged over the area to improve the signal-to-noise ratio. By comparing the measured signals with the FEA results of fused silica (relative permittivity $\varepsilon_r = 3.8$) and LaAlO$_3$ ($\varepsilon_r \sim 25$), we can estimate a radius of curvature $r \sim 600$ nm at the apex. With this fitting parameter determined, the MIM-Im AC signals can be simulated as a function of permittivity of isotropic dielectrics in Fig. 3. The experimental data on 8 bulk materials are also plotted in the graph, showing quantitative agreement with the modeling results. For each anisotropic dielectric, we calculate an effective isotropic permittivity by FEA simulation such that the MIM AC response matches that using the actual anisotropic permittivity tensor. Note that for anisotropic samples, the measured value is closer to the permittivity perpendicular to the sample surface, e.g., along the c-axis for (001) surface. Such a phenomenon can be explained by the monopole-like tip geometry, which generates quasi-static electric fields mostly in the vertical direction.

Fig. 4a shows the topographic and MIM-Im AC images of the Al/SiO$_2$/Si sample measured by the same tip as above. The microwave image is clearly dominated by the electrical response since the insulating surface contamination particle (~ 100 nm in height) and conductive Al dots (~ 20 nm in height) exhibit opposite contrast with respect to the substrate. In traditional contact-mode MIM, only the contrast between the Al dots (covered by 4 ~ 5 nm native oxide) and the 100-nm-$SiO_2$/Si background is meaningful. In TF-based MIM, however, the absolute signals at both regions (~ 40 mV on Al and ~ 8 mV on the substrate) represent local electrical properties and can be readily simulated by the FEA, as shown in Fig. 4b. An additional advantage of distance modulation at the kHz range is that the signal does not suffer from the electronic thermal drift in the time scale of minutes[30]. As a result, no background removal is needed to post-process the MIM-Im AC raw data in Fig. 4a. The spatial resolution of ~ 300 nm, as inferred from the line profile across one Al dot (Fig. 4c), is consistent with the tip diameter. Owing to the robustness of the DAM operation, high-quality AFM and MIM imaging can be acquired at a fast scan rate (up to 10 µm/s) without obvious scan instability and tip wear.

Finally, we demonstrate that the TF-based MIM is capable of performing quantitative conductivity imaging on nano-devices, which is one of the key areas of application in microwave microscopy. Fig. 5a shows the optical and AFM images of an exfoliated $MoS_2$ FET device on $SiO_2$/Si substrate. Details of the device structure and analysis of the contact-mode MIM results can be found in Ref. 25. Here the carrier density in the $MoS_2$ flake can be globally tuned by the back-gate voltage $V_{BG}$. In Fig. 5b, selected MIM-Im/Re AC images at various $V_{BG}$'s are displayed and substantial local inhomogeneity is observed in the channel region. To compare the macroscopic transport and microscopic imaging results, we plot the transfer characteristics (source-drain conductance $G_{DS}$ versus $V_{BG}$) in Fig. 5c and MIM-Im/Re AC signals as function of $G_{DS}$ over an area of 1.2 µm × 1.2 µm in Fig. 5d. Using the same FEA process above, we can also simulate the MIM-Im/Re AC signals as a function of the sheet conductance $g_{sh} = \sigma \cdot h$, where $\sigma$ is the conductivity and $h$ is the thickness of $MoS_2$. The response curves in Fig. 5e are similar to that of the contact-mode MIM[9], except that the signals are now absolute values rather than relative contrast over the insulating background. In particular, the MIM-Im AC signals increase monotonically as $g_{sh}$ increases and saturates at both the insulating ($g_{sh} < 10^{-9}$ S·sq) and conductive ($g_{sh} > 10^{-5}$ S·sq) limits. The MIM-Re AC signals, on the other hand, peak at an intermediate $g_{sh} \sim 10^{-7}$ S·sq. Comparing the data in Fig. 5d and simulation in Fig. 5e, the local sheet conductance within the dashed square in the MIM images can be quantitatively extracted. The results in Fig. 5f nicely track the transport behavior in Fig. 5c, with deviations due to the strong inhomogeneity in the sample and contact resistance in the device. The error bar indicates the uncertainty of conductivity measurement which, as shown in the inset of Fig. 5f, can be larger close to the insulating and conductive limits as a result of the saturated MIM AC signal. We emphasize that the spatial variation of $g_{sh}$ carries rich information on the material properties and device performance. The ability to quantitatively map out the conductance distribution is therefore highly desirable for fundamental and applied research.

## IV. Conclusions

In summary, we have demonstrated quantitative measurements using tuning fork-based microwave impedance microscopy operated in the driving amplitude modulation mode. The demodulated MIM AC signal can be simulated by a combination of FEA and Fourier transformation. Excellent agreement is achieved between the modeling and the experiment data on both bulk dielectrics and working nano-devices. Our work provides the pathway to perform quantitative near-field microwave imaging, where absolute signal levels can be readily interpreted as the local permittivity and conductivity.

## Acknowledgements

This research is funded by the U.S. Department of Energy (DOE), Office of Science, Basic Energy Sciences, under the Award No. DE-SC0010308. The authors thank E.Y. Ma and Y.T. Cui for helpful discussions.

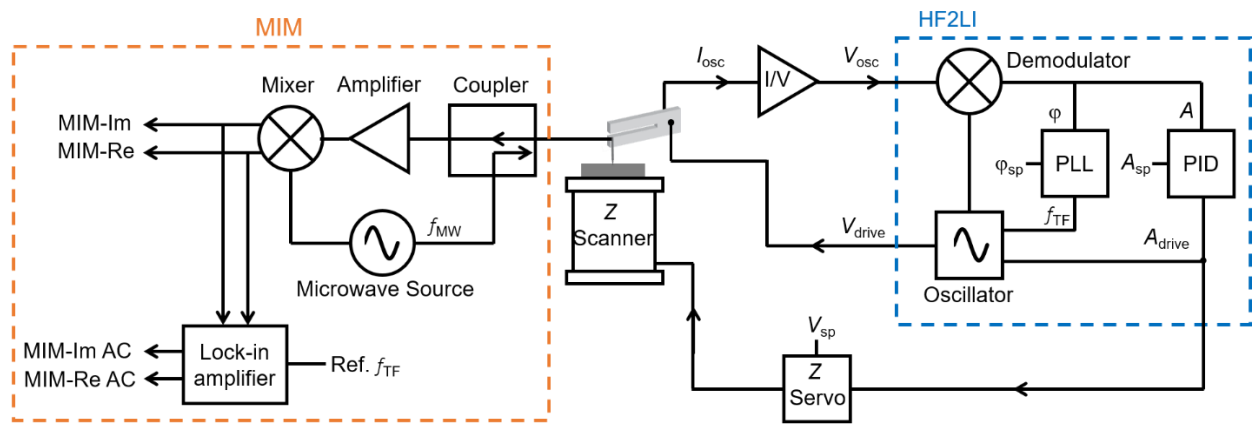

FIG. 1. Schematic of the TF-based AFM configured for the DAM mode and the microwave electronics (detailed in the text).

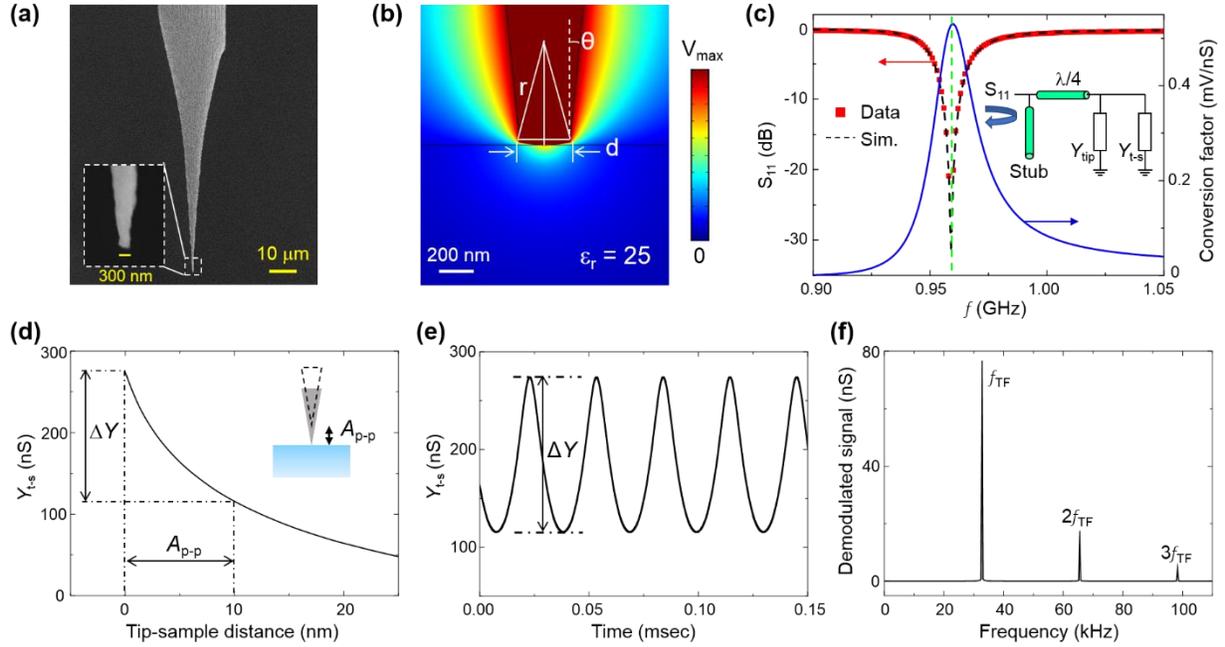

FIG. 2. (a) SEM image of a typical etched W tip. The inset shows a zoom-in view near the tip apex. (b) Quasi-static potential distribution around the tip and a bulk dielectric sample simulated by the FEA software. (c) Measured $S_{11}$ (red squares) of a TF-based sensor and a fit to the transmission line analysis (black dashed line). The blue curve is the simulated conversion factor between the tip-sample admittance and the MIM output. The inset shows the equivalent circuit of the impedance-match network. (d) Tip-sample admittance $Y_{t\text{-}s}$ on a dielectric sample ($\varepsilon_r = 25$) as a function of tip-sample distance (sketched in the inset) modeled by FEA. The tip oscillates between the contact point and a maximum height of $A_{p\text{-}p}$, resulting in a change of $\Delta Y$ in the tip-sample admittance. (e) Time dependence of $Y_{t\text{-}s}$, assuming a simple harmonic oscillation of the tip. (f) Fourier spectrum of $Y_{t\text{-}s}$. The amplitude of the first harmonic peak corresponds to the MIM AC signal demodulated by the lock-in amplifier.
11

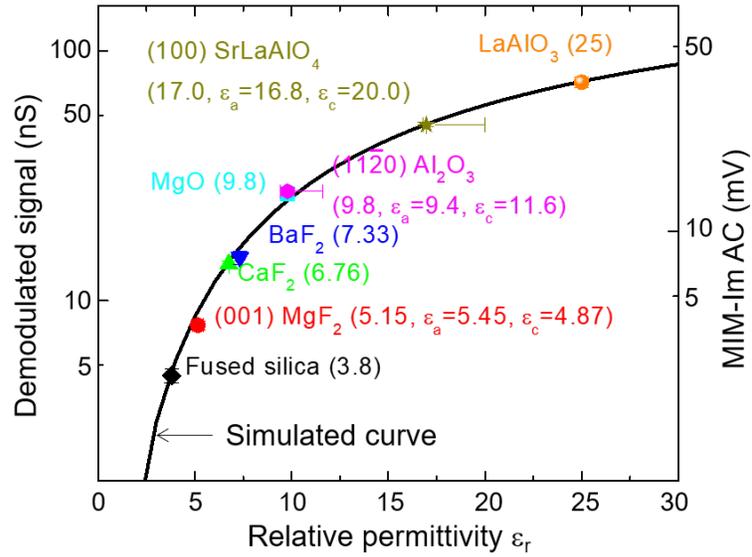

FIG. 3. Demodulated tip-sample admittance and the corresponding MIM-Im AC signals as a function of the relative permittivity. Materials and the permittivity values are listed next to the symbols. Bars in the vertical axis indicate experimental uncertainties. Bars in the horizontal axis, on the other hand, identify the range between permittivity values at a-axis ($\varepsilon_a$) and c-axis ($\varepsilon_c$) for anisotropic materials. Data points for anisotropic materials represent the effective isotropic permittivity (see the text for details). The solid line is the simulated curve using FEA.

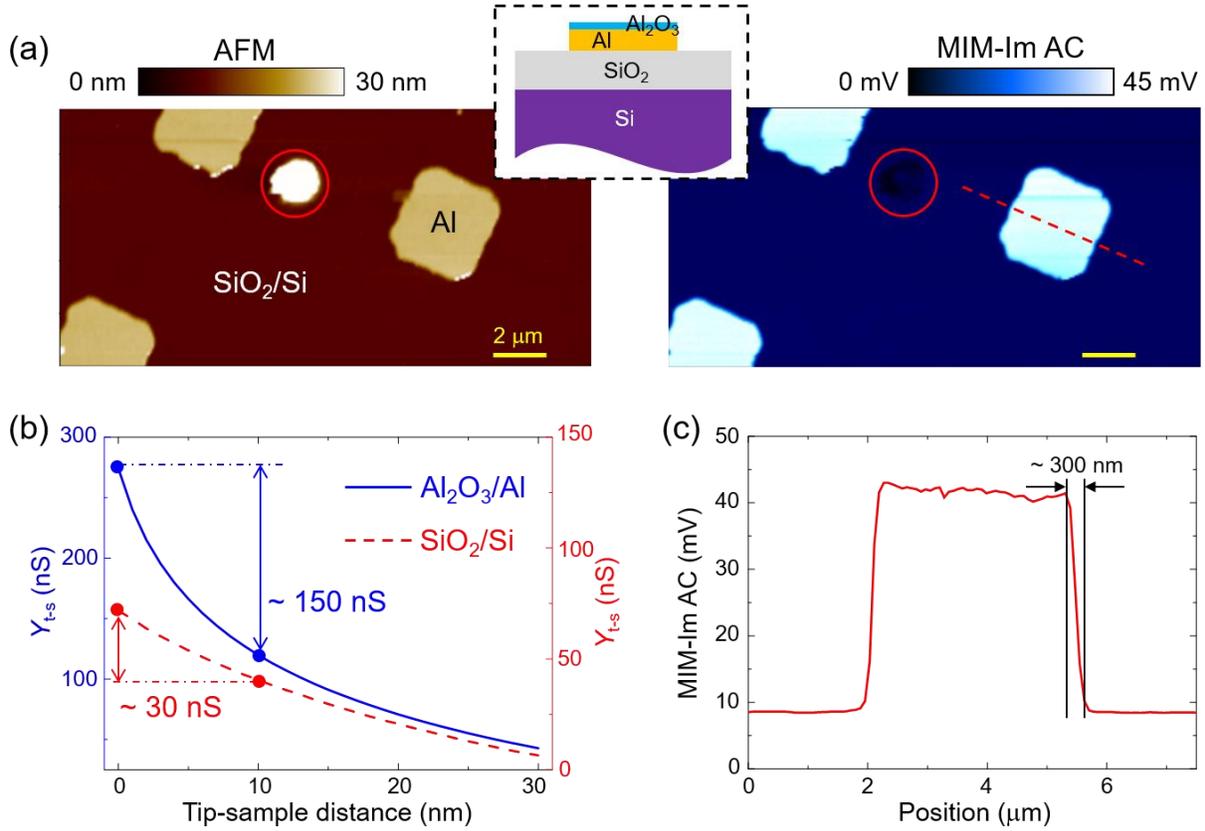

FIG. 4. (a) AFM and MIM-Im AC images of a patterned Al dot sample measured by the TF-based MIM. The scale bars are 2 µm. The inset shows the sample structure. An insulating surface particle (inside the red circle) shows higher topographic and lower MIM signals than the substrate. (b) Simulated approach curves on the Al dot (solid blue line) and the substrate (dashed red line). (c) Line profile of the MIM-Im AC signal across an Al dot marked in (a), showing the electrical contrast between the two regions and a spatial resolution of ~ 300 nm.

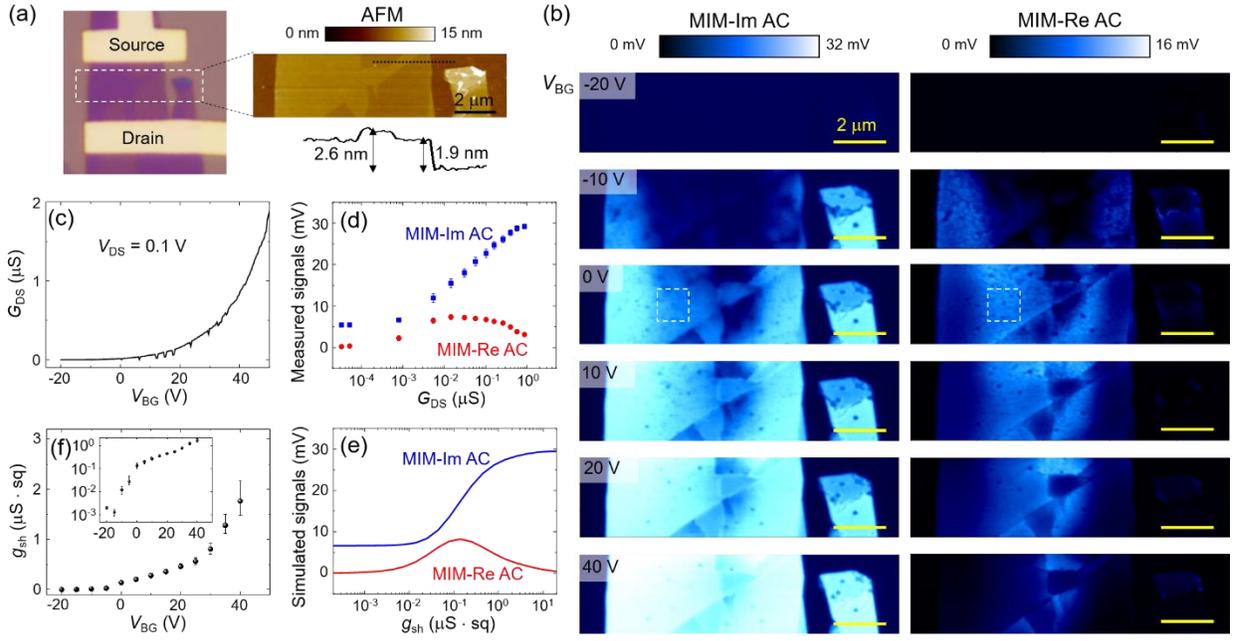

FIG. 5. (a) Optical and AFM images of a back-gated $MoS_2$ field effect transistor. (b) Selected MIM-Im/Re AC images at different back gate voltages. All scale bars are 2 µm. (c) Transfer characteristics of the device at a source-drain voltage $V_{DS}$ = 0.1 V. (d) Averaged MIM AC signals inside the dashed square in (b) as a function of source-drain conductance $G_{DS}$. (e) Simulated MIM AC signals as a function of the sheet conductance $g_{sh}$. (f) Local sheet conductance versus $V_{BG}$ calculated by comparing (d) and (e). The inset shows the same data with y-axis in log scale.